\documentclass{iau} 
\usepackage{graphicx}

\title[JD 11.~~GWs from Flaring Magnetars] 
{Search for gravitational-wave transients associated with magnetar bursts during the third Advanced
LIGO and Advanced Virgo observing run}

\author[Kara Merfeld on behalf of the LIGO, Virgo, and KAGRA collaborations]   
{Kara Merfeld$^1$ on behalf of the LIGO, Virgo, and KAGRA collaborations}

\affiliation{$^1$University of Oregon \\ Department of Physics, Eugene, OR 97403, USA \\ email: {kmerfeld@uoregon.edu} \\[\affilskip]}

\pubyear{2021}
\volume{363}  
\jname{Neutron Star Astrophysics at the Crossroads: Magnetars and the Multimessenger Revolution}
\editors{A.C. Editor, B.D. Editor \& C.E. Editor, eds.}
\begin{document}

\maketitle

\begin{abstract}
Magnetars are neutron stars with exceptionally strong dipole magnetic fields which are observed to display a range of x-ray flaring behavior, but the flaring mechanism is not well understood. 
The third observing run of Advanced LIGO and Virgo extended from April 1, 2019 to March 27, 2020, and contained x-ray flares from known magnetar SGR 1935+2154, as well as the newly-discovered magnetar, Swift J1818-1607. 
We search for gravitational waves coincident with these magnetar flares with minimally modeled, coherent searches which specifically target both short-duration gravitational waves produced by excited f-modes in the magnetar's core, as well as long-duration gravitational waves motivated by the Quasi-Periodic Oscillations observed in the tails of giant flares. 
In this paper, we report on the methods and sensitivity estimates of these searches, and the astrophysical implications.
\end{abstract}

\firstsection 
\section{Introduction}

Magnetars - neutron stars with exceptionally strong external dipole magnetic fields- display a range of bursting activity including hard x-ray flares and soft gamma ray flares with peak luminosities typically $\lesssim10^{43}\,{\rm erg\,s}^{-1}$, as reviewed by \cite[Kaspi and Beloborodov (2017)]{Kaspi_2017}, and giant flares that have been observed to have peak fluences up to $10^{48}\,{\rm erg\,s}^{-1}$ (\cite{evans80,hurley99,hurley05,boggs07}).
Quasi-Periodic Oscillations (QPOs) have been observed in the hundreds of seconds following some giant flares (\cite{barat83,israel05,strohmayer05,strohmayer06,watts06}), with frequencies ranging between ~$18$ Hz and $2384$ Hz.  
These QPOs might couple to torsional modes in the crust, and while it is unlikely that these torsional modes themselves would produce observable gravitational waves (GWs), they might couple to and excite f-modes in the magnetar's core.  
In this paper we present the search methodology and sensitivity estimates of a search for GWs coincident with magnetar x-ray bursts during the third observing run of Advanced LIGO and Virgo.



\section{O3 Magnetar bursts}

Our present analysis methods require strain data from at least 2 observatories in order to conduct a coherent search, and there were 15 galactic magnetar bursts detected by Fermi GBM and reported in the \cite{IPN} at times during O3 when this condition was satisfied. 
One additional burst is considered in our study, as Virgo entered observing mode 83s after the burst time and is still able to provide enough data in which to search for long-duration gravitational waves.  
Eleven of these bursts came from SGR $1935+2154$, a galactic magnetar of particular interest due to its association with the fast radio burst on April 28th, 2020 (\cite{CHIME}), and also its recently discovered periodic windowed behavior (\cite{Grossan2021}). 
Two of these bursts are from newly-discovered magnetar Swift J$1818-1607$. 
Three of these bursts came from an unidentified source, but are assumed to have been produced by the same source because they occur within an 18 hour window on February 3rd, 2020.  
All three detections had very poor sky localization,
and we search for GWs from these flares assuming that they were produced by the closest of the magnetars in the 3$\sigma$ error region of all of the flares, 1 RXS J170849 at a distance of 3.8kps (\cite{Durant2006}), so that we can obtain more meaningful upper limits.  
The information on each burst in the study is contained in Table~\ref{tab1}.

\begin{table}
  \begin{center}
  \caption{The magnetar bursts included in this study, taken from interplanetary network \cite{IPN}.  
  These are the flares that took place during O3 when at least 2 detectors were in observing.  
  The detectors in the network are denoted as H (LIGO Hanford), L (LIGO Livingston) and V (Virgo). 
  The isotropic electromagnetic energy is calculated using the fluences provided in \cite{lin2020fermi}. 
  Bursts 2669-2671 are from an unknown source, as reported by \cite{von2020fourth}.}
  \label{tab1}
 {\scriptsize
  \begin{tabular}{|l|c|c|c|c|c|c|c|}\hline 
    {\bf Burst} & {\bf Source} & {\bf Date} &  {\bf Time } &  {\bf Detectors} & { \bf E$_\mathrm{EM}^\mathrm{iso}$ } & {\bf Reference} \\  
   &  & & [UTC] & & [ergs] & \\ \hline
    2651 & SGR 1935 & Nov 04, 2019 & 01:54:37 & H V* & - & GCN 26169\\
    2652 & SGR 1935 & Nov 04, 2019 & 02:53:31 & H V & $1.43 \times 10^{39}$ & GCN 26163, 26151\\
    2653 & SGR 1935 & Nov 04, 2019 & 04:26:55 & H L V & $1.14 \times 10^{39}$ & GCN 26163\\
    2654 & SGR 1935 & Nov 04, 2019 & 06:34:00 & H L V & $ - $ & GCN 26153\\
    2655 & SGR 1935 & Nov 04, 2019 & 09:17:53 & H L & $5.72 \times 10^{39}$  & GCN 26163, 26154\\
    2656 & SGR 1935 & Nov 04, 2019 & 10:44:26 & H L & $ 2.23 \times 10^{40}$ & GCN 26242, 26163, 26158, 26157\\
    2657 & SGR 1935 & Nov 04, 2019 & 12:38:38 & H L V & $2.74 \times 10^{39} $ & GCN 26163 \\
    2660 & SGR 1935 & Nov 04, 2019 & 15:36:47 & H V & $ 1.20 \times 10^{39} $ & Fermi GBM catalog \\
    2661 & SGR 1935 & Nov 04, 2019 & 20:29:39 & H V & $1.33 \times 10^{39}$ & GCN 26165, 26166\\
    2665 & SGR 1935 & Nov 05, 2019 & 06:11:08 & H V & $7.81 \times 10^{40}$ & GCN 26242\\
    2668 & SGR 1935 & Nov 15, 2019 & 20:48:41 & L V & $7.68 \times 10^{38}$ & Fermi GBM catalog\\
  \hline
    2669 & - & Feb 03, 2020 & 03:17:11 & H L V & - & GCN 26980\\
    2670 & - & Feb 03, 2020 & 03:44:03 & H L V & - & GCN 26969, 26980\\
    2671 & - & Feb 03, 2020 & 20:39:37 & H L V & - & GCN 26980\\
  \hline
    2673 & Swift J1818 & Feb 28, 2020 & 22:19:32 & L V & - &Fermi GBM catalog \\
    2674 & Swift J1818 & Mar 12, 2020 & 21:16:47 & H L V & - & GCN 27373\\
  \hline

  \end{tabular}
  }
 \end{center}
\vspace{1mm}
 \scriptsize{
 {\it Notes:}\\
  $^1$ $*$ denotes a detector which was out of observing mode at the time of the flare, but which came into observing and still provides enough data to be included. \\
}
\end{table}

\section{Search Methods}
We use two searches for short duration (ms to s in duration) and long duration (100s of s) gravitational waves.  The short-duration search is designed to be most sensitive to GWs that are produced by excited f-modes in the neutron star's core, while the long duration search is conducted after the flare time, and is sensitive to GWs in the same frequency range as the QPOs.  Both of the searches are minimally modeled, which reflects the uncertainty of the underlying physics.

{\underline{\it Short Duration Search:}}
The short duration search is designed to have optimal sensitivity to the properties of a GW produced by an f-mode.  We use X-Pipeline (\cite{Sutton2010,Was2012}), a GW data analysis package, to conduct a coherent search on data from multiple detectors.  The data around the time of the flare is combined into a time-frequency map, displaying the whitened energy per pixel.  The brightest 1\% of pixels are then selected out and assigned an SNR.  Then neighboring pixels are combined into clusters and their SNRs are summed to provide an SNR of the cluster.  These clusters then undergo consistency tests, and the SNRs of the surviving clusters are compared to the SNRs of the surviving clusters in the data taken from before and after the burst, which comprise the background.  This is a targeted search over the time and location of each flare.

We conduct 2 different short-duration searches which we call the \textit{centered search} and the \textit{delayed search}.  The centered search is centered on the time of the burst and searches a window spanning [-4s,4s] around the time of the flare.  The purpose of including this search over a shorter span of time is to optimize our sensitivity right at the time with the highest emission of electromagnetic energy, and when the GW is most likely to be emitted.  The frequency range is from 50Hz-4000Hz, to include the entire f-mode range, and to give sensitivity at lower frequencies as well. 

The delayed search is conducted on data [+4s, +504s] after the burst time, and on the same frequency range as the centered search. We include this search so that we have sensitivity to short-duration signals, specifically from excited f-modes, that might be coincident with the QPOs observed in the x-ray tails of the giant flares.  It is important to note that none of the bursts included in our study are giant flares, and none of them have observed QPOs.  But if the x-ray bursts in our study are produced by a diminished version of the same internal mechanism as the giant flares, then they might still have excited f-modes after the burst time.

{\underline{\it Long Duration}}
The search for long duration gravitational waves is conducted using the  Stochastic Transient Analysis Multi-detector Pipeline (STAMP) \cite{Thrane2011}, which coherently combines the data from 2 detectors into a time-frequency map, and each pixel on the map is assigned an SNR.  Bezier curves varying in frequency by less than 10\% are then randomly generated, and the SNR of a Bezier curve is the sum of the SNRs of the pixels it crosses through.  The SNR of the gravitational wave candidate at the time of the burst is the SNR of the loudest Bezier curve overlapping that time-frequency map.  A false alarm probability is then generated by comparing this SNR to the analogous highest SNRs taken from times both before and after the time of the burst, which comprise the background.  We search in the frequency band that QPOs have been observed in, $25$Hz$-2500$Hz, and our search window is after the time of the burst, from [+4s, +1604s].

\section{Sensitivity Estimates}
The sensitivities of the short and long duration searches are determined by injecting simulated GW signals into the data.  The pipelines then determine the root sum squared injection amplitude ($h_{rss,50}$) at which the injection is correctly identified with 50\% efficiency, where:

$$  h_{rss} = \sqrt{ \int _{-\infty}^\infty \left| h(t) \right| ^2 dt },$$
$$h(t) = \left|h_+(t)\right|^2 + \left|h_\times(t)\right|^2.$$

The $h_{rss,50}$ is then used to calculate a gravitational-wave energy sensitivity at which $50$\% of signals would have been detected, $E_{50}$.  

The injected waveforms are chosen such that their morphology most closely mimics the properties of a gravitational wave.  
For both components of the short-duration search, we model f-modes using both unpolarized sine Gaussians and ringdowns at central frequencies ranging from $70$Hz$-3560$Hz. 
In order to better constrain some models, we also inject circularly polarized sine Gaussians at 1600Hz and 2020Hz.  
We also include 4 White Noise Burst injections, characterized by frequency independent amplitude, in frequency ranges $100$Hz$-200$Hz, and $100$Hz$-1000$Hz, and durations $11$ms and $100$ms. 
The injections for the long duration search include half sine Gaussian and ringdown waveforms at frequencies ranging from $55$Hz$-1550$Hz, with damping times $150$s and $400$s.

The search sensitivity depends on the detector characteristics at the times of the bursts, and on the orientation of the Earth-based detectors.  
For a 145Hz sine Gaussian in the short duration search, we approximate the $h_{rss,50}$ as being $6\times 10^{-23}$Hz$^{-\frac{1}{2}}$, and for a $150$Hz half sine Gaussian in the long-duration search, we approximate $h_{rss,50}$ as $ 1\times 10^{-22}$Hz$^{-\frac{1}{2}}$.  
The short duration search has slightly worse sensitivity at higher frequencies, with $h_{rss,50}$ approximately $ 1\times 10^{-22}Hz^{-\frac{1}{2}}$.

\section{Conclusions}
A framework for the magnetar burst search in the third LIGO/Virgo/KAGRA run is presented.  
We are sensitive to short-duration signals (ms to s) in the f-mode frequency range, and to low-frequency long duration signals which have $h_{rss,50}$ approximately $1\times 10^{-22}Hz^{-\frac{1}{2}}$.  
The full results of the study are available at \cite{LVK2022}.

SGR 1935+2154 has shown periodic windowed behavior, so we should expect to see more bursts from it during the 4th observation run, which can be searched in a similar study to the one presented here.  It will also be interesting to conduct a stacked analysis of the SGR 1935+2154 events from O3, and those from O4.  Acknowledgments for this work may be found in https://dcc.ligo.org/LIGO-P2100218/public.

\end{document}